\documentclass{article}
\usepackage{spconf,amsmath,graphicx}
\usepackage{cite}
\usepackage{amsmath,amssymb,amsfonts}
\usepackage{algorithm, algorithmic}
\usepackage{graphicx}
\usepackage{textcomp}
\usepackage{array}
\usepackage{tabularx}
\usepackage{makecell}
\usepackage{booktabs}
\usepackage{multicol}
\usepackage{multirow}
\usepackage{xcolor}
\usepackage{kotex}

\usepackage{enumitem}

\title{Language-Oriented Communication with Semantic Coding and Knowledge Distillation for Text-to-Image Generation}
\name{$^\dagger$Hyelin Nam, $^\ddagger$Jihong Park, $^\ddagger$Jinho Choi, $^*$Mehdi Bennis, and $^\dagger$Seong-Lyun Kim\thanks{This work was supported in part by the Institute of Information \& communications Technology Planning \& Evaluation (IITP) grant (No. 2021-0-00347) funded by the Ministry of Science and ICT (MSIT), and in part by the Information Technology Research Center (ITRC) support program (IITP-2023-RS-2023-00259991) supervised by IITP. J. Park and S.-L. Kim are corresponding authors (email: jihong.park@deakin.edu.au, slkim@yonsei.ac.kr).}}
\address{$^\dagger$Yonsei University, $^\ddagger$Deakin University, and $^*$University of Oulu}
\begin{document}
%
\maketitle
\begin{abstract} 
By integrating recent advances in large language models (LLMs) and generative models into the emerging semantic communication (SC) paradigm, in this article we put forward to a novel framework of language-oriented semantic communication (LSC). In LSC, machines communicate using human language messages that can be interpreted and manipulated via natural language processing (NLP) techniques for SC efficiency. To demonstrate LSC’s potential, we introduce three innovative algorithms: 1) semantic source coding (SSC) which compresses a text prompt into its key head words capturing the prompt's syntactic essence while maintaining their appearance order to keep the prompt's context; 2) semantic channel coding (SCC) that improves robustness against errors by substituting head words with their lenghthier synonyms; and 3) semantic knowledge distillation (SKD) that produces listener-customized prompts via in-context learning the listener's language style. In a communication task for progressive text-to-image generation, the proposed methods achieve higher perceptual similarities with fewer transmissions while enhancing robustness in noisy communication channels.

\end{abstract}
\begin{keywords}
Semantic communication (SC), large language model (LLM), generative model.
\end{keywords}
\section{Introduction}
\label{sec:intro}

Semantic communication (SC) is an emerging research paradigm that focuses on the meanings (i.e., semantics) and effectiveness of communicating bits \cite{gunduz2022beyond,qin2022semantic,barbarossa2023semantic,seo2023towards,10054510}. Deep joint source and channel coding (DeepJSCC) is a prime example wherein an encoder-decoder structured neural network (NN) acts as a transceiver, within which task-effective features are extracted from input data and made into communication messages. These \emph{neural messages} are the NN's hidden-layer activations trained and tailored for a specific task, which greatly improves communication efficiency \cite{gunduz2022beyond,qin2022semantic,barbarossa2023semantic,seo2023towards}. 


However, neural messages constraint the full potential of SC. First, NN activations are not always universal messages, as they are influenced by their training data and communication environment. Indeed, DeepJSCC transceivers that have been trained separately are hardly interoperable without fine-tuning \cite{choi2023semantics}. Furthermore, the semantics of these NN activations are nothing but what remains after achieving effective communication. It is therefore difficult to interpret and manipulate these semantics as intended.




\begin{figure}
    \centering
    \includegraphics[width=\columnwidth]{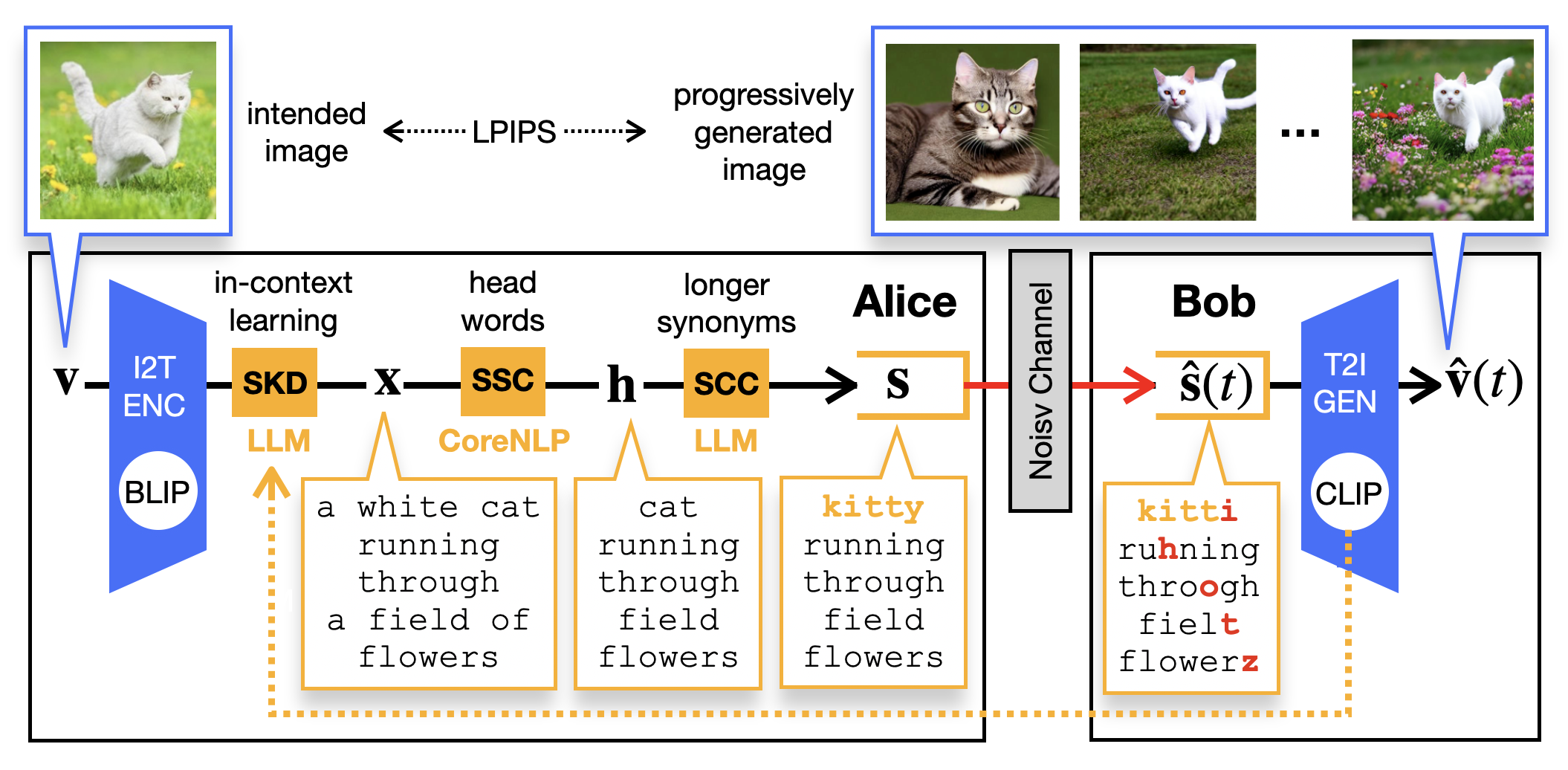}
    \caption{An illustration of language-oriented semantic communication (LSC) for progressive text-to-image generation, empowered by semantic source coding (SSC), semantic channel coding (SCC), and semantic knowledge distillation (SKD).}
    \label{fig: systemmodel}
\end{figure}

Human language, by contrast, is universal and versatile to describe a broad range of tasks, owing to its evolution throughout diverse human experiences in history. Moreover, with recent advances in natural language processing (NLP) and generative models, machines are now capable of interpreting and manipulating human language. Motivated by this, in this paper we propose a novel \emph{language-oriented SC (LSC)} framework, which facilitates SC through human \emph{language messages}. The operation of LSC transceivers is threefold:
\begin{enumerate}[leftmargin=.5cm]
    \item[1)] \textbf{Data-to-Language Translation}: Text-based cross-modal models transform input data into language messages to be transmitted (e.g., via CLIP for image-to-text (I2T) or Whisper for speech-to-text translation).

    \item[2)] \textbf{Language Analysis \& Manipulation}: Large language models (LLMs) and other NLP algorithms (e.g., GPT-4\cite{openai2023gpt4}, Llama 2\cite{touvron2023llama}, and CoreNLP \cite{manning-EtAl:2014:P14-5}) are utilized for analyzing the syntax, semantics, and context in language messages and manipulating these messages for improving communication efficiency.
    
    \item[3)] \textbf{Language-to-Data Generation}: Text-conditioned generative models produce intended data using the received message seman (e.g., via Stable Diffusion for text-to-image (T2I) or Zeroscope for text-to-video generation).

\end{enumerate}


To showcase the potential of LSC, this paper we consider a point-to-point LSC scenario, where Alice sends a text prompt describing an intended image, word by word, while Bob progressively generates an image based on the accumulated received text prompt. The LSC accuracy between Alice and Bob is assessed using the learned perceptual image patch similarity (LPIPS) that measures the distance between intended and generated images. To improve the communication efficiency of LSC, as visualized in Fig.~\ref{fig: systemmodel}, we focus on Step 2), and develop the following novel algorithms:
\begin{itemize}[leftmargin=.3cm]
    \item \textbf{Head-based Semantic Source Coding (SSC)} is a lossy compression of the original prompt by pruning non-head words, inspired from our empirical findings that sending all words in a prompt does not always achieve the lowest LPIPS. The heads of a prompt are the words determining the syntactic category of the prompt, which can be identified, for example, using CoreNLP \cite{manning-EtAl:2014:P14-5, chen2014fast}.

    
    \item \textbf{Synonym-based Semantic Channel Coding (SCC)} adds redundancy into the prompt by replacing original head words with their longer synonyms, increasing the robustness to channel noise perturbing each character of the words. Only the synonyms ensuring the same semantics of the prompt are of interest, which can be found by, for instance, using GPT-4 \cite{openai2023gpt4}.

    \item \textbf{In-Context Learning-based Semantic Knowledge Distillation (SKD)} aims to address the out-of-distribution (OOD) prompts due to different language knowledge between Alice and Bob, and enables Alice to emulate Bob's prompts by assimilating Bob's language knowledge. This can be achieved without re-training NN model parameters, by harnessing LLM's unique capability of in-context learning, i.e., few-shot learning via demonstration \cite{brown2020language}.


\end{itemize}

Simulation results reveal SSC compresses transmitted messages by up to $42.6$\%, while surprisingly reducing LPIPS by $0.015$ compared to full prompts. Applying SCC and SKD further cuts LPIPS by up to $0.007$ and $0.009$ by addressing channel noise and heterogeneous language knowledge.

\vspace{5pt}
\noindent\textbf{Related Works}: Recent research \cite{barbarossa2023semantic} employs generator models in SC but with neural messages, distinct from LSC's language messages. The ts\_zip \cite{tszip} algorithm exploits LLM-based synonyms for compression, differing from our synonym-based SCC for robustness and from our head-based SSC. LSC stands apart from other language-based SC studies that mainly focus on I2T compression or on the entropy of text truthfulness \cite{carnap1952outline}, in contrast to LSC harnessing LLM and NLP techniques to analyze and manipulate language messages for SC.

\begin{table*}[t]
\caption[c]{Message compression ratios of SSC with SSC and/or SKD.}
    \centering
    \resizebox{0.9\textwidth}{!}{
    \begin{tabular}{lccc|ccc|ccc}
         \toprule
          Compression& \textbf{w.o.SSC} & \textbf{SSC} & \textbf{SSC+SKD} & \textbf{SSC} & \textbf{SSC+SCC} & \textbf{SSC+SCC+SKD} & \textbf{SSC} & \textbf{SSC+SCC} & \textbf{SSC+SCC+SKD}  \\
          Ratio & \multicolumn{3}{c|}{w.o. channel noise} & \multicolumn{3}{c|}{SNR = $6.25$ dB} & \multicolumn{3}{c}{SNR = $8.75$ dB} \\
        \hline
        w.r.t. Word & $1$ & \textbf{0.641} & $0.654$ & \textbf{0.641} & \textbf{0.641} & \textbf{0.641} & \textbf{0.654}& \textbf{0.654} & \textbf{0.654} \\
        
        w.r.t. Character & $1$ & \textbf{0.426} & $0.468$ & \textbf{0.426} & $0.531$ & $0.525$& \textbf{0.468} & $0.579$ & $0.565$ \\
        \hline
        LPIPS & $0.718$ & $0.703$ & \textbf{0.697}& $0.736$ & $0.730$ & \textbf{0.721} & $0.726$ & $0.719$ & \textbf{0.715} \\
        \bottomrule
        
    \end{tabular}}
    \label{table: cr}
\end{table*}

\section{Semantic Source Coding for Progressive Text-to-Image Generation} \label{section: SSC}

In this section, we propose SSC for a progressive text-to-image (T2I) generation task in a point-to-point communication scenario, as elaborated next. 

\vspace{5pt}
\noindent  1) \textbf{Image-to-Text Translation}:\quad Alice has an intended image $\mathbf{v}$ to send, and translates it into a text prompt $\mathbf{x}$, a sequence containing a set $\mathbf{X}$ of words presented in a specific order, given as:
\begin{align}
\mathbf{x}=\texttt{I2T}(\mathbf{v}) = ( \mathbf{x}_1, \mathbf{x}_2,\cdots, \mathbf{x}_{|\mathbf{X}|} ),
\end{align}
where $\mathbf{x}_i$ is the $i$th word comprising $|\mathbf{x}|$ characters. The function $\texttt{I2T}(\cdot)$ represents an image-to-text (I2T) encoder such as BLIP \cite{li2022blip} or CLIP \cite{radford2021learning}.

\vspace{5pt}
\noindent 2) \textbf{Head-based Semantic Source Coding (SSC)}: Alice aims to compress and transmit text characters of $\mathbf{x}$ while maintaining the semantics of $\mathbf{x}$. The semantics can be maintained when the key words of $\mathbf{x}$ are presented without loosing their syntax and context. To this end, SSC first identifies a set $\mathbf{H}$ of $\mathbf{x}$'s head words that determine the prompt's syntactic category in linguistic analysis. While keeping head words' order of appearance in $\mathbf{x}$, SSC produces a compressed sequence $\mathbf{h}  = (\mathbf{h}_1, \mathbf{h}_2, \cdots, \mathbf{h}_{|\mathbf{X}|}) $ in which:
\begin{align}
\mathbf{h}_t = 
    \begin{cases}
    \mathbf{x}_i, & \text{if } \mathbf{x}_i \in  \mathbf{H}\\
    \emptyset,              & \text{otherwise}.
    \end{cases}
\end{align}
The head words in $\mathbf{H}$ can be identified using the CoreNLP algorithm \cite{chen2014fast}, i.e., $\mathbf{H} = \texttt{CoreNLP}(\mathbf{x})$. 
Consequently, SSC yields the compression ratio $|\mathbf{H}|/|\mathbf{X}| \leq 1$ in terms of words, and $\sum_{\mathbf{x}_i \in \mathbf{H}} |\mathbf{x}_i|/\sum_{\mathbf{x}_i\in \mathbf{X}} |\mathbf{x}_i |$ in terms of characters.


\vspace{5pt}
\noindent 3) \textbf{Text-to-Image Generation}:
Bob receives the head words of $\mathbf{h}_i$ in order, and progressively generates an image using a T2I generator such as Stable Diffusion \cite{rombach2022high} and DALL$\cdot$E. At the $i$-th head word reception with $i\in \{1, 2, \cdots, |\mathbf{H}|\}$, the received prompt is $\mathbf{h}(t)$, and the generated image is:
\begin{align}
\hat{s}(t) = \texttt{T2I}(\mathbf{h}(t)) = \texttt{T2I}( (\mathbf{h}_1, \mathbf{h}_2, \cdots, \mathbf{h}_t) ). 
\end{align}
The perceptual similarity between Bob's generated $\mathbf{\hat{v}}(t)$ and Alice's intended image $\mathbf{v}$ is measured by the learned perceptual image patch similariy (LPIPS) score \cite{zhang2018unreasonable} that calculates the distance at hidden layers of pre-trained AlexNet, given as:
\begin{align}
\texttt{LPIPS}(\mathbf{v}, \mathbf{\hat{v}}(t)) = \sum_{l} \frac{1}{H_l W_l} \sum_{h,w} \Vert f(\mathbf{v})-f(\mathbf{\hat{v}}(t) \Vert^{2}_{2}.
\end{align}
The term $l$ identifies the $l$-th layer having its width $w$, height $h$, and dimension $H_l\times W_l$ with an activation function~$f(\cdot)$.


\section{Semantic Channel Coding in Noisy Communication Channels}

In the previous section, we presume that Alice's transmitted head words are perfectly received at Bob. In this section, we consider a noisy channel, and propose SCC to address noisy head word receptions at Bob.

\vspace{5pt}
\noindent 1) \textbf{Noisy Channel Model}: Alice individually transmits a set $C_{\mathbf{h}_t}$ of characters in the head word $\mathbf{h}_t$. Following a discrete memoryless channel (DMC) model, Bob receives the head word $\mathbf{\hat{h}}_t$ containing a set $\hat{C}_{\mathbf{h}_t}$ of characters, each of which is perturbed as a different character with a cross-over probability $\epsilon>0$ and is otherwise successfully received. 

\vspace{5pt}
\noindent 2) \textbf{Synonym-based Semantic Channel Coding (SCC)}: In this noisy channel, SCC aims to enhance $\mathbf{h}_t$'s robustness by increasing $|\mathbf{C}_{\mathbf{h}_t}|$ while maintaining the same semantics of the prompt $\mathbf{h}(t)$. In the aforementioned channel, Bob encounters the same error in each characters following a geometric distribution. This does not allow communicating short words like "$\texttt{cat}$" that changes its semantics even with a single-character variation (e.g., $\texttt{bat}$, $\texttt{cut}$, and $\texttt{car}$), motivating SCC. In SCC, we consider a set $\mathbf{S}_{\mathbf{h}_t}$ of candidate synonyms of $\mathbf{h}_t$, given as:
\begin{align}
\mathbf{S}_{\mathbf{h}_t} = \{ \mathbf{s}_1, \mathbf{s}_2, ... \mathbf{s}_{|\mathbf{S}_{\mathbf{h}_t}|}  \},
\end{align}
where $\mathbf{s}_t$ contains a set $\mathbf{C}_{\mathbf{s}_t}$ of characters. Although $\mathbf{S}_{\mathbf{h}_t}$ can be found using a dictionary, it ignores the context of $\mathbf{h}(t)$,and does not guarantee the intended semantics. To solve this, SCC utilizes an LLM such as GPT-4 and Llama 2, a decoder-only autoregressive model that can predict the most in-context appropriate synonym $\mathbf{s}_t^*$ of $\mathbf{h}_t$ in $\mathbf{h}(t)$ by masking $\mathbf{h}_t$ and maximizing the following conditional unmasking probability:
\begin{align}
\mathbf{s}_t^* = \max_{\mathbf{s}_j \in \mathcal{W}}  p_(\mathbf{s}_j)= \Pr( \mathbf{s}_j | \mathbf{h}(t)\backslash \mathbf{h}_t),
\end{align}
where $\mathcal{W}$ is a set of total characters, e.g., $128$ characters in ASCII.
By relaxing this LLM, we obtain a set $\mathbf{\hat{S}}_{\mathbf{h}_t}$ of in-context synonyms associated with their unmasking probabilities exceeding a threshold $p_c >0$:
\begin{align}
\mathbf{\hat{S}}_{\mathbf{h}_t} =  \{\mathbf{\hat{s}}_1, \mathbf{\hat{s}}_2, \cdots,  \mathbf{\hat{s}}_{.} \} = \{ \mathbf{s}_j | p_{\mathbf{s}_j} \geq p_c \}.
\end{align}
Consequently, SSC can increase noise robustness of $\mathbf{h}_t$ within a set $L_{\mathbf{h}_t}$ of the levels in terms of characters, given as
\begin{align}
L_{\mathbf{h}_t} = \{ | \mathbf{\hat{s}}_j | \in \mathbf{\hat{S}}_{\mathbf{h}_t} |  |\mathbf{h}_t|  \leq  |\mathbf{\hat{s}}_j |   \leq L_c \},
\end{align}
where $L_c$ is the number of characters of the lengthiest synonym of $\mathbf{h}_t$, i.e., $L_c = \lfloor \max_{\mathbf{\hat{s}}_j\in \mathbf{\hat{S}}_{\mathbf{h}_t}} | \mathbf{C}_{\mathbf{\hat{s}}_j}| \rfloor$.

\section{Semantic Knowledge Distillation in Heterogeneous Language Knowledge}

In this section, we aim to address the problem when Alice and Bob have different knowledge on text-image relations by proposing SKD that enables Alice to produce Bob-customized text prompts via in-context learning.

\vspace{5pt}
\noindent 1) \textbf{Heterogeneous Knowledge Model}: 
BLIP and CLIP are encoder-decoder NN models that store knowledge on image-text relations through cross-attention weights, enabling T2I and I2T conversions. Suppose that Alice has BLIP for T2I while Bob utilizes CLIP for I2T. This incurs OOD generation in both, decreasing LPIPS. 

\vspace{5pt}
\noindent 2) \textbf{In-Context Learning-based Semantic Knowledge Distillation (SKD)}: 
A pre-trained LLM has excessive knowledge containing spurious correlations, and conditioning its knowledge within a specific context can improve task performance. In-context learning enables this by feeding few exemplary input-output pairs to teach the LLM a desired context, i.e., few-shot learning via demonstration \cite{brown2020language}. Meanwhile, knowledge distillation (KD) is a method to transfer a target (teacher) model's knowledge into another (student) model by minimizing their output differences for common inputs \cite{hinton2015distilling}. Inspired from in-context learning and KD, SKD shows $K$ input images to Alice and Bob that generate $K$ output text prompts, using BLIP and CLIP, respectively, that are fed into an LLM for demonstration. This in-context learned LLM becomes a text-to-text (T2T) translator that can produce Bob-customized prompt $\hat{\mathbf{x}}_b$ for a given Alice's prompt $\mathbf{x}_a$, given~as:
\begin{align}
\hat{\mathbf{x}}_b = \texttt{T2T}\left(\mathbf{x}_a; \{ \mathbf{v}^{(i)}, \mathbf{x}_a^{(i)}, \mathbf{x}_b^{(i)} \}_{i=1}^K \right),
\end{align}
where $\mathbf{x}_a^{(i)}$ and $\mathbf{x}_b^{(i)}$ are the $i$-th output prompts at Alice and Bob, respectively. After SKD, $\hat{\mathbf{x}}_b$ is fed into SSC and/or SCC. Note that SKD is applicable before SSC and/or after SCC. We focus on the former for simplicity.
\section{Numerical Results}

\begin{figure}[t]
    \centering
    \includegraphics[clip,width=\columnwidth]{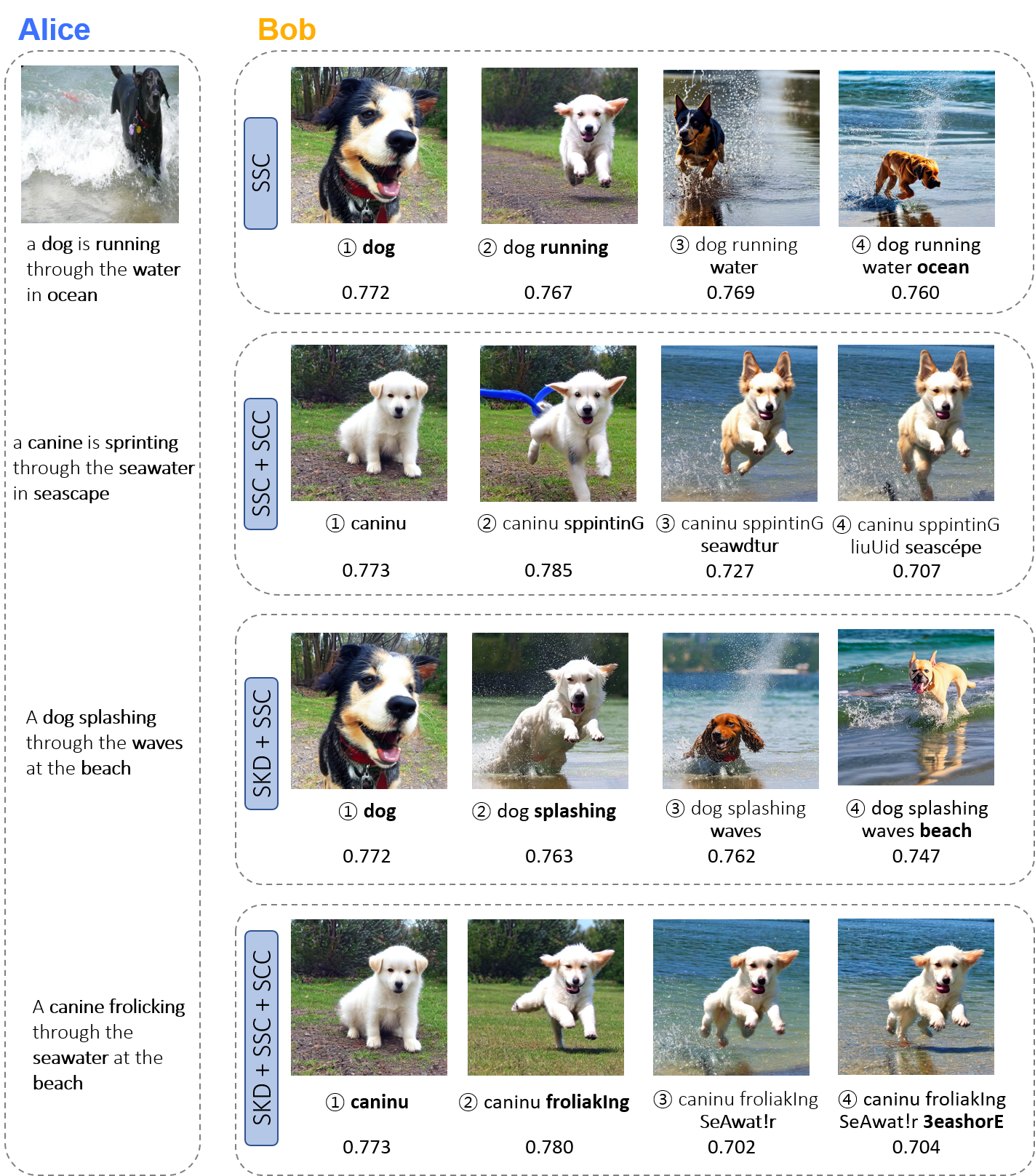}
    \caption{Alice's image and prompts (left) and Bob's generated images and LPIPS (right), with SSC, SCC, and~SKD.}
    \label{fig: sample}
\end{figure}

\noindent\textbf{Simulation Settings}:
We consider that Alice's I2T encoder is BLIP \cite{https://doi.org/10.48550/arxiv.2201.12086}, and Bob's T2I generator is Stable Diffusion v1.5 \cite{rombach2022high} that generates each image from a prompt with $50$ denoising steps. This diffusion process is conditioned by the text prompt encoded using CLIP \cite{radford2021learning}. SKD and SCC that require LLMs are based on GPT-4 \cite{openai2023gpt4}, while SSC is run by CoreNLP \cite{manning-EtAl:2014:P14-5, chen2014fast}. For image data, the Flicker8k dataset is considered, containing 8,092 samples, each with 256x256 pixels \cite{Flickr8k}. For text prompts, each character is 8-bit ASCII coded, and modulated using 16QAM. During SCC, $p_c$ is set as $0.72$. All LPIPS values are averaged over 100 simulation runs, except for snapshot visualizations in Fig.~\ref{fig: sample}.

\begin{figure}[t]
    \centering
    \includegraphics[clip,width=.92\columnwidth]{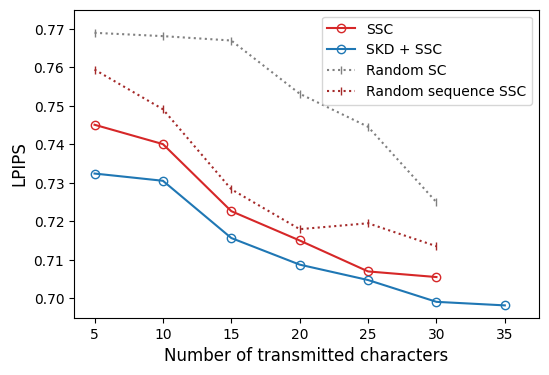} \vspace{-15pt}
    \caption{SSC with or without SKD w.r.t. transmitted characters.}
    \label{fig: char}
\end{figure}

\begin{figure}[t]
    \centering
    \vspace{-10pt}
    \includegraphics[clip,width=.92\columnwidth]{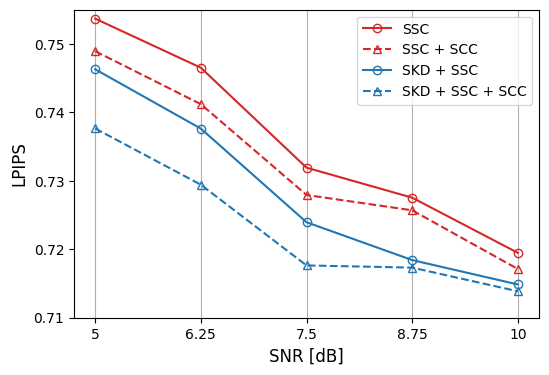}
    \vspace{-15pt}
    \caption{SSC with or without SCC and SKD w.r.t. SNR.}
    \label{fig: snr}
\end{figure}


\vspace{5pt}
\noindent\textbf{Impact of SSC}: 
Tab. \ref{table: cr} reveals SSC achieves 64.1\% word compression and 42.6\% in character. Surprisingly, mean LPIPS improves by $0.015$, suggesting SSC's roles not only in compression but also in prompt engineering. Fig. \ref{fig: char} highlights SSC's dual benefits (solid red), head extraction and appearance-based ordering. To dissect their LPIPS reduction contributions, we introduce two baselines: \emph{Random SC} (dotted gray), which maintains appearance order with the same compression ratio as SSC but transmits random words from $\mathbf{x}$ instead of head words; and \emph{Random sequence SSC} (dotted red), sending head words from $\mathbf{h_t}$ like SSC but in a shuffled order. Results indicate head extraction reduces mean LPIPS by up to $0.04$, and appearance-based ordering contributes up to a $0.012$ reduction, as seen when comparing SSC with Random SC and Random sequence SSC, respectively.



\vspace{5pt}
\noindent\textbf{Impact of SCC}: 
In Fig.~\ref{fig: snr}, we observe that mean LPIPS decreases with SNR. Comparing SSC+SCC (dotted red) to SSC (solid red), SCC contributes to a reduction in mean LPIPS by up to $0.007$. This reduction diminishes with SNR. However, SCC compromises compression, increasing it by up to 10.5\% as shown in Tab.~\ref{table: cr}. In these simulations the increase in characters is capped at $4$. In certain instances, as illustrated in Fig.~\ref{fig: sample} at $\text{SNR}=7.5$dB, the LPIPS reduction from SCC can be as much as 7.57x its average. This suggests potential benefits in optimizing SSC level based on given channel conditions for future research.


\vspace{5pt}
\noindent\textbf{Impact of SKD}: 
As illustrated in Figs.~\ref{fig: char} and \ref{fig: snr}, SKD contributes to a reduction in mean LPIPS by up to $0.006$ and $0.009$, respectively. Notably, the latter reduction surpasses even the contribution of SCC to LPIPS reduction. However, SKD may extend the prompt length, e.g., by an average of $5$ characters in Fig.~\ref{fig: char}, highlighting a trade-off between compression and LPIPS.




\vspace{-5pt}
\section{Conclusion}
In this article we proposed LSC, and developed SSC, SCC, and SKD that leverage NLP and LLM techniques to improve LSC's SC efficiency under noisy channels and heterogeneous T2I/I2T knowledge. Future research might explore various tasks such as I2T-based control and compare LSC's performance with its DeepJSCC counterpart. It could also be interesting to exploit other LLM capabilities such as reasoning and interactions with humans.


\vfill \newpage \clearpage


\vfill\pagebreak

\bibliographystyle{IEEEbib} 

\end{document}